\documentclass[twocolumn,secnumarabic,showpacs,preprintnumbers,amssymb,floatfix,prb]{revtex4}
\usepackage{amsmath,amssymb}
\usepackage{epsfig,pstricks}

\begin{document}

\title{Electron transport in crossed nanotubes with point contact}

\author{Viktor Margulis and Mikhail Pyataev}

\email{pyataevma@math.mrsu.ru}

\affiliation{Mordovian State University, Saransk, Russia}.

\date{\today}

\begin{abstract}
The electron transport in a four-terminal nanodevice consisting of two crossed
nanotubes is investigated in the framework of the Landauer-Buttiker formalism.
The evident formula for the ballistic conductance of the device
is found using a model of crossed conductive cylinders
with a point contact between them.
Sharp conductance dips stipulated by resonance scattering on the contact
are shown to appear in the conductance of the first cylinder.
The conductance between the cylinders has resonant behavior.
The form and the position of resonant peaks are studied.
Our results indicate that the form of asymmetric dips and peaks in the conductance
differs from the well-known Fano line shape. We have shown that the maximal
value of the conductance between cylinders does not exceed
a unit of the conductance quantum.
\end{abstract}


\maketitle

\section{Introduction}

Carbon nanotubes are considered as a promising material for future
nanoelectronic engineering due to their unique physical properties.
The quantum coherent transport properties of carbon nanotubes have been
confirmed by experimental results, indicating that the tube can be considered
as a ballistic conductor at least up to 200 nm. \cite{Lia01}
Phase coherent transport and electron interference have been observed
in single-walled \cite{Kon01} and multi-walled \cite{Zha06} carbon nanotubes.
It was shown that superlattice properties of
carbon nanotubes in a transverse electric field raise
new possibilities for developing optoelectronic devices operating
in the terahertz range of frequencies \cite{KPP05}.
Recent theoretical studies have shown that electronic devices
together with their metallic interconnects can, in principle, be fabricated
on a single tube. \cite{Kil00}
Furthermore, nanotubes can operate under a variety of
conditions and are compatible with many other materials
and fabrication techniques.
Several promising functional devices based on carbon nanotubes
have been proposed. \cite{Ana06}
These devices include metallic wires, \cite{Tan97}
terahertz range emitters \cite{Kib05}
field effect transistors, \cite{Tan98,Mar98,Pos01,Bac01,App04}
and nanometer-size rectifying diodes. \cite{Yao99,Fuh00,Zho00,Lee04}

The design of the integrated electronic circuits implies the application
of many contacted nanotubes.
Therefore, transport properties of the contacts between nanotubes are
of particular interest. In the last few years, electron transport
in such contacts has been investigated experimentally. \cite{Fuh00,Kim03,Gao04,Hei05}

A number of interesting theoretical models
has been suggested to study physical properties
of the contacts. \cite{Chi03,Men04,Dag04,Kim05,Kim07,Yan05,Val05}
The transmission through atom-contacted single-walled carbon nanotubes
were calculated within the tight-binding approach. \cite{Chi03}
Multiterminal junctions of single-walled carbon nanotubes were investigated
using the classical molecular dynamics method.  \cite{Men04}
Tight-binding calculations of the conductance of multiply connected
metallic carbon nanotubes were carried out in Refs. \onlinecite{Kim05,Kim07}.
Parallel and crossed junctions of single-wall carbon nanotubes
were studied in the framework of the tight-binding approximation. \cite{Dag04}
Transport properties of three-terminal carbon
nanotube junctions have been investigated
within the scattering matrix approach. \cite{Val05}
The differential conductance of several crossed carbon nanotubes were
calculated using the tight-binding model and the Green's function method. \cite{Yan05}

It has been shown that the conductance through several
multiterminal nanotube junctions exhibits Fano resonances.
\cite{Kim05,Kim07,Val05,Yan05}
This phenomenon emerges from the coherent interaction
of a discrete state and a continuum and was first discovered
by studying the asymmetric peak in helium spectrum. \cite{Fan61}
Subsequent theoretical investigations have shown
the occurrence of this effect in numerous mesoscopic devices, including
quantum dots \cite{Noc92,CWB01} and
quasi-one-dimensional channels with impurities. \cite{KS99, KS99PRB, KS99E, KSR02}
Recently the resonances were observed experimentally in the ballistic conductance
of a single-electron transistor \cite{GGH00}
and a quantum ring. \cite{KAK02}
Asymmetric dips and peaks similar to the Fano resonances
have been found in the conductance of carbon nanotubes. \cite{Kim03,Bab04,Zha06}
The interest to this problem is stipulated by the possibility of application
the phenomenon in high-sensitive resonant electronic devices.
The resonances lead to  large changes in current intensity
in short intervals of voltage. This phenomenon may be used in designing
of precise electronic devices.

The necessary condition for the emergence of Fano resonances is the existence
of the discrete level in the continuous spectrum. In the case of carbon nanotube
junctions, these discrete levels were attributed to pentagonal and heptagonal defects
of the honeycomb lattice. \cite{Kim05,Kim07,Val05,Yan05}
However, the detailed atomic structure of the multiterminal nanotube junction
still was not studied experimentally.
Therefore, the origin of asymmetric line shapes
in the conductance requires further theoretical investigations,
especially in the case of multiwall carbon nanotubes.

It should be noted that the most of theoretical studies
of the electron transport in the junctions
were focused on nanotubes of sufficiently small diameters (about 1-4 nm).
At the same time, the diameters of tubes used in some experiments \cite{Kim03}
were in the range of 25-30 nm. The application of the tight-binding approach
to this systems is somewhat difficult because it requires considerable
amount of computer resources. Furthermore, physical meaning of the phenomenon
is sometimes smeared in sufficiently accurate but very complicated models.
Thus, it should be useful to study the electron transport through the contact
between two nanotubes using a simple model which allows exact analytical solution.
The simplest model with the geometry of a nanotube is a structureless
two-dimensional cylindrical surface.
Transport properties of the electron gas
on the cylindrical surface have already been studied in the literature.
\cite{Cha98,Mag98,MST00,Nem06,Gri07}
In particular, this model has been used for analysis of
some electron properties of carbon nanotubes. \cite{Nem06,Gri07}
Other interesting systems with cylindrical geometry
are rolled GaAs/AlGaAs heterostructures. \cite{Pri00,Vor04}

\section{Hamiltonian}

\begin{figure}[htb]
\includegraphics[width=0.75\linewidth]{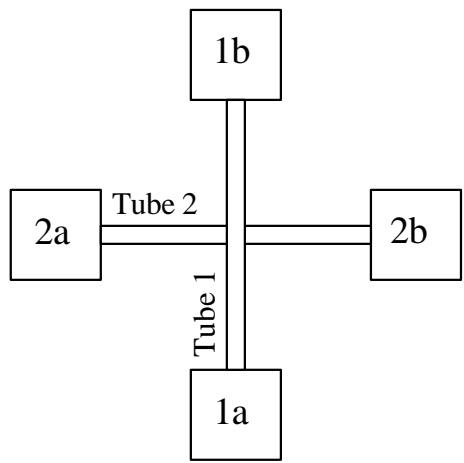}
\caption{\label{scheme}
Scheme of the device. Tube 1 lies over the tube 2.
Rectangles 1a,1b,2a, and 2b, represent electron reservoirs.}
\end{figure}

The purpose of the present paper is the theoretical investigation
of the electron transport in a four-terminal nanodevice consisting
of two crossed nanotubes with a point contact between them.
Each tube is modeled by a conductive cylindrical surface of radius $r_j$ $(j=1,2)$.
The schematic view of the device is shown in Fig.~\ref{scheme}.

Our first goal is to construct the electron Hamiltonian of the system.
If we ignore the contact between the cylinders, then the electronic states
are described by the unperturbed Hamiltonian
$H_0=H_1\oplus H_2$,
where $H_1$ and $H_2$ are electron Hamiltonians in the first and the second cylinder
respectively. In this case, the electron wave function may be represented
in the form of one-column matrix
\begin{equation}
\psi=\left(\begin{array}{c}
\psi_1\\
\psi_2
\end{array}
\right).
\end{equation}
Since the contact between cylinders is modeled by a single point,
the Hamiltonian $H$ of the whole system is a point perturbation
of the operator $H_0$.
To obtain this perturbation we use the zero-range potential theory.

We introduce two independent cylindrical coordinate systems and
denote the point on the cylinder by
${\bf q}=(z,\varphi)$.
Then the Hamiltonian $H_j$ has the form
\begin{equation}
H_j=\frac{p_z^2}{2m_j}+\frac{L_z^2}{2m_j r_j^2},
\end{equation}
where  $m_j$ is the electron effective mass in $j$-th nanotube,
$p_z$ and  $L_z$ are projections of the momentum and the angular momentum
onto the axis of the cylinder.
The electron effective mass depends on the size and chirality of the tube.
For each cylinder, we use its own cylindrical coordinate system.

The spectrum of the Hamiltonian $H_j$ is given by
the sum of the discrete part $E_m^{(j)}=\varepsilon_{j} m^2$
and the continuous part  ${p_z^2}/({2 m_j})$
\begin{equation}
E_{m,p_z}^{(j)}=\varepsilon_{j} m^2+{p_z^2}/({2 m_j}),
\end{equation}
where $m$ is the magnetic quantum number
and $\varepsilon_j=\hbar^2/(2m_j r_j^2)$.
In the case of identical tubes ($r_1=r_2$), we will omit below the subscript
of $\varepsilon$.

To obtain the Hamiltonian $H$ of the whole system
we have to define the point perturbation of the Hamiltonian $H_0$.
For this purpose, we use linear boundary conditions
at the point of the contact. Boundary values for the wave function $\psi_j$
of the electron in $j$-th cylinder are determined with the help
of the zero-range potential theory. \cite{DO88,Alb88,Bru03}
The theory shows that the electron wave function
$\psi_j({\bf q})$ has the logarithmic singularity
in a vicinity of the contact point ${\bf q}_0$
\begin{equation}
\label{asymp}
\psi_j({\bf q})=-u_j \ln \rho({\bf q},{\bf q}_0)+v_j+R({\bf q}),
\end{equation}
where $\rho({\bf q},{\bf q}_0)$ is geodesic distance between the points ${\bf q}$ and ${\bf q}_0$,
$u_j$ and $v_j$ are complex coefficients, and  $R({\bf q})\to 0$
in the limit  ${\bf q} \to {\bf q}_0$. The similar method has been
used earlier in Refs.~\onlinecite{GMP03,MP05}.

It is clear that the boundary conditions at the point of contact
are some linear relations between $u_1$, $v_1$, $u_2$, and $v_2$.
The coefficients of the relations are not all
independent \cite{Bru03} since the Hamiltonian $H$ is Hermitian.
Thus, the most general form of the boundary conditions is given by
\begin{equation}
\label{bound}
\begin{cases}
v_1-b_1 u_1=a u_2,\\
v_2-b_2 u_2=a^* u_1.
\end{cases}
\end{equation}
Here the coefficients $b_1$ and $b_2$ determine the strength of the
zero-range potential at the point of contact
and $a$ is a dimensionless parameter that is
responsible for the coupling of the wave functions on
different cylinders.
According to the zero-range potential theory parameters $b_j$
can be represented in terms of scattering lengths $\lambda_j$
by the relation $b_j=2 \ln \lambda_j$.
It should be noted that the zero-range potential is attractive and
the strength of the potential decreases with
increasing of $\lambda_j$. The limit $\lambda_j \to \infty$ corresponds
to the absence of the point perturbation.
We point out that the model of zero range potential is applicable
when the size of the contact is much smaller
than the Fermi wavelength of the electron.

\section{Transmission coefficients}
In the paper, we investigate the conductance $G_{11}$ of the first cylinder
and the conductance $G_{21}$ that is responsible
for the electron transport from the first cylinder to the second one.
According to the Landauer--B\"uttiker formula the zero-temperature conductance  $G_{ji}$
can be expressed in terms of transmission coefficients $T^{ji}_{m'm}$
from the state with magnetic quantum number $m$ in $i$-th cylinder to the state
with $m'$ in $j$-th cylinder
\begin{equation}
\label{Land}
G_{ji}=G_0\sum\limits_{m'm} T^{ji}_{m'm}.
\end{equation}
Here $G_0=e^2/\pi\hbar$ is the conductance quantum
and the sum is taken over all states with $E_m\leq \mu$, where $\mu$ is the Fermi energy.

The transmission coefficients are represented via transmission amplitudes $t_{m'm}^{ji}$
$$
T_{m'm}^{ji}=\frac{k_{m'}^{(j)}}{k_m^{(i)}}|t_{m'm}^{ji}|^2.
$$
To determine the amplitudes $t_{m'm}^{ji}$ we need a solution
of the Schr\"odinger equation for the Hamiltonian $H$.
The zero-range potential theory allows us to
represent the solution in terms of the
Green function $G_j({\bf q},{\bf q}',E)$ for the operator $H_j$
\begin{equation}
\label{psi}
\begin{cases}
\psi_1({\bf q})=\psi_m({\bf q})+\alpha_1 G_1({\bf q},{\bf q}_0;E),\\
\psi_2({\bf q})=\alpha_2 G_2({\bf q},{\bf q}_0;E),
\end{cases}
\end{equation}
where $\psi_m({\bf q})=\exp( ik_m z+im\varphi)$ is an incident wave, and
coefficients $\alpha_j$ have to be determined from boundary conditions.
The Green function $G_j({\bf q},{\bf q}',E)$ is given by \cite{MP05}
\begin{equation}
\label{G1}
G_j({\bf q}, {\bf q}'; E)=\frac{i m_j}{2 \pi \hbar^2}
\sum\limits_{m=-\infty}^{\infty}
\frac{e^{ik_m^{(j)}|z-z'|+im(\varphi-\varphi')}}{k_m^{(j)}r},
\end{equation}
where $k_m^{(j)}=\sqrt{2m_j(E-E_m^{(j)})}/\hbar$,
$\mathop{\rm Re} k_m \geq 0$, and $\mathop{\rm Im} k_m \geq 0$.

Substituting the wave function (\ref{psi}) into Eq.~(\ref{asymp}),
we obtain
\begin{equation}
\begin{cases}
u_j=\frac{i m_j}{2 \pi \hbar^2}\alpha_j,\\
v_j=\delta_{j1}+\alpha_j Q_j(E),
\end{cases}
\end{equation}
where $Q_j(E)$ is Krein's Q-function \cite{Bru03}
that is the renormalized Green function of the Hamiltonian $H_j$
\begin{displaymath}
Q_j(E)=\displaystyle\lim\limits_{{\bf q} \to {\bf q}_0}
\left[\frac{2\pi\hbar^2}{m_j}G_j({\bf q}_0 ,{\bf q} ;E)+ \ln \rho({\bf q}_0,{\bf q})
\right].
\end{displaymath}
To simplify our notations we denote $\widetilde Q_j(E)=Q_j(E)-b_j$.
The explicit form of $\widetilde Q_j(E)$ was found in Ref. \onlinecite{MP05}
\begin{equation}
\label{eq:Qj}
\widetilde Q_j(E)=\frac{i}{r_j k_0^{(j)}}+2\sum\limits_{m=1}^{\infty}
\left(\frac{i}{r_j k_m^{(j)}}-\frac{1}{m} \right)+2\ln\frac{r_j}{\lambda_j}.
\end{equation}

One can see from Eqs. (\ref{psi}) and  (\ref{G1})
that the wave function $\psi_j({\bf q})$
has the following asymptotics at $z \to \infty$:
\begin{equation}
\psi_j({\bf q})\simeq
\sum\limits_{m'=m_{\rm min}}^{m_{\rm max}}
t_{m'm}^{ji} e^{ik_{m'}^{(j)} z+i m'\varphi},
\end{equation}
where $m_{\rm min}$ and $m_{\rm max}$  are minimal
and maximal values of the magnetic quantum number for occupied states.
The asymptotics of the wave function $\psi_1({\bf q})$
in the first cylinder at $z \to -\infty$ is given by
\begin{equation}
\nonumber
\psi_1({\bf q})\simeq
e^{ik_{m}^{(1)} z+i m\varphi}+
\sum\limits_{m'=m_{\rm min}}^{m_{\rm max}}
r_{m'm}^{11} e^{-ik_{m'}^{(1)} z+i m'\varphi},
\end{equation}
where $r_{m'm}^{11}$ are reflection amplitudes.
Elementary but cumbersome calculations show that the relation
\begin{equation}
\nonumber
\sum\limits_{m'=m_{\rm min}}^{m_{\rm max}}
\left[
\frac{k_{m'}^{(1)}}{k_{m}^{(1)}}(|r_{m'm}^{11}|^2+|t_{m'm}^{11}|^2)
+2\frac{k_{m'}^{(2)}}{k_{m}^{(1)}}|t_{m'm}^{21}|^2\right]=1
\end{equation}
is valid for an arbitrary energy $E$ that is the manifestation of
the current conservation law for our system. Here factor $2$ corresponds to equal
probabilities for an electron to pass from lead $1a$ to lead $2a$ or $2b$
(see  Fig.~\ref{scheme}).

Applying boundary conditions (\ref{bound}) to the wave function (\ref{psi}),
we obtain the following form for the transmission amplitudes $t_{m'm}^{11}$:
\begin{equation}
\label{tmm}
t_{m'm}^{11}(E)=\delta_{mm'}-\frac{i\widetilde Q_2}{\widetilde Q_1 \widetilde Q_2-|a|^2}.
\end{equation}
The transmission coefficients $T_{m'm}^{11}$ are given by
\begin{equation}
\label{T11}
T_{m'm}^{11}(E)=\frac{k_{m'}^{(1)}}{k_{m}^{(1)}}
\left|\delta_{mm'}-\frac{i\widetilde Q_2}{\widetilde Q_1\widetilde Q_2-|a|^2}\right|^2.
\end{equation}

We note that according to Eq.~(\ref{eq:Qj})
\begin{equation}
\label{eq:ImQj}
\mathop{\rm Im} \widetilde Q_j=\sum\limits_{m=m_{\rm min}}^{m_{\rm max}}\frac{1}{r_j k_m^{(j)}}.
\end{equation}
Thus, we can express the sum in Landauer's formula (\ref{Land}) in terms of the Q-function
and obtain the following equation for the conductance $G_{11}(\mu)$:
\begin{equation}
\label{G11}
\frac{G_{11}(\mu)}{G_0}= N(\mu)-\frac{(\mathop{\rm Im} \widetilde Q_1)^2|\widetilde Q_2|^2}{|\widetilde Q_1 \widetilde Q_2-|a|^2|^2}-
2\frac{|a|^2 \mathop{\rm Im} \widetilde Q_1\mathop{\rm Im} \widetilde Q_2}{|\widetilde Q_1 \widetilde Q_2-|a|^2|^2}.
\end{equation}
Here $N(\mu)=m_{\rm max}-m_{\rm min}+1$ is the number of states with the energy
smaller than $\mu$.
It is worth mentioning that the possibility to represent the conductance
in the explicit form is based on the application of zero-range potentials
for modeling the contacts.

\section{Results and discussion}
\begin{figure}[htb]
\includegraphics[width=0.95\linewidth]{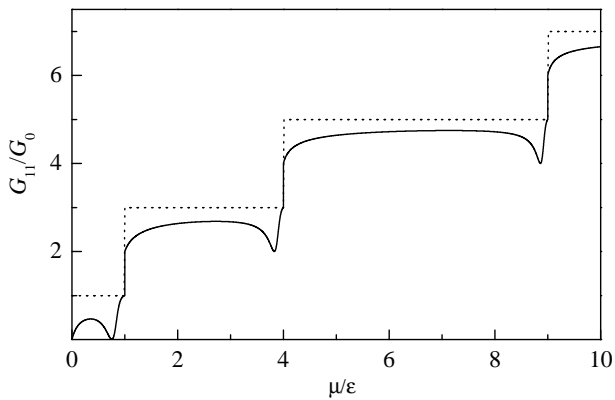}
\caption{ \label{f:G11-1}
Conductance  $G_{11}$ versus the Fermi energy $\mu$
at $r_1=r_2$,  $\lambda_1=\lambda_2=3r_1$, and  $a=0.1$.
Dotted line represents the conductance of the unperturbed cylinder.}
\end{figure}

\begin{figure}[htb]
\includegraphics[width=0.95\linewidth]{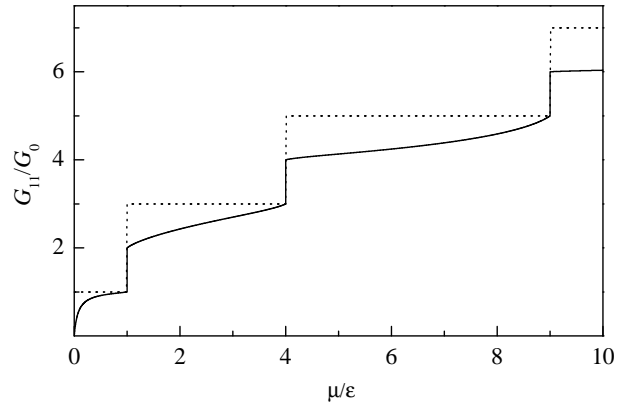}
\caption{\label{f:G11-2}
Conductance  $G_{11}$ versus the Fermi energy $\mu$
at $r_1=r_2$, $\lambda_1=\lambda_2=0.15r_1$, and  $a=0.1$.
As in the previous figure,
dotted line represents the conductance of the unperturbed cylinder.}
\end{figure}

\begin{figure}[htb]
\includegraphics[width=0.95\linewidth]{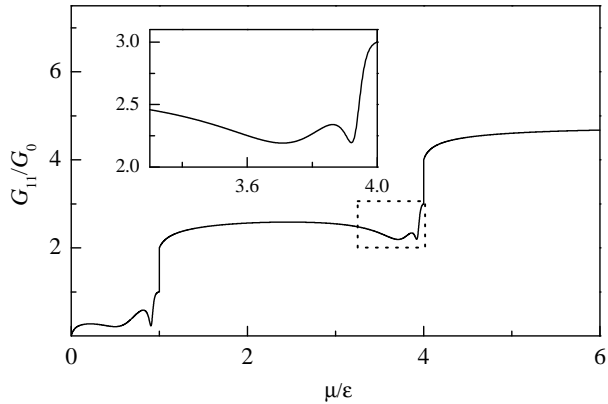}
\caption{\label{f:G11-3}
Conductance  $G_{11}$ versus the Fermi energy $\mu$
at $r_1=r_2$,  $\lambda_1=\lambda_2=4r_1$, and $a=4$.
The inset represents a fragment of the curve marked with the dotted rectangle.}
\end{figure}

\begin{figure}[thb]
\includegraphics[width=0.95\linewidth]{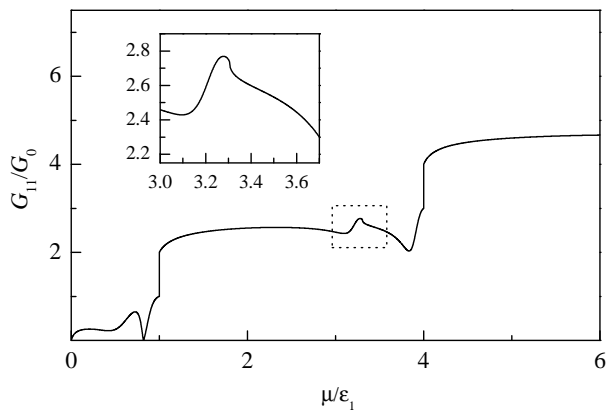}
\caption{\label{f:G11-4}
Conductance  $G_{11}$ versus the Fermi energy $\mu$
at $r_2=1.1 r_1$,  $\lambda_j=4r_j$, and  $a=4$.
The inset represents a fragment of the curve marked with the dotted rectangle.}
\end{figure}

Conductance $G_{11}$ as a function of the Fermi energy $\mu$ is represented in
Figs. \ref{f:G11-1}-\ref{f:G11-4}.
If the contact between the cylinders is absent, then Eq.~(\ref{G11})
contains only the first term, and the dependence $G_{11}(\mu)$ is step-like.
The second term in Eq.~(\ref{G11}) is responsible for the back-scattering
on the contact point and the last term is stipulated by
the transmission of electrons from the first nanotube to the second one.

At first, we consider the case of identical tubes ($r_1=r_2$ and $\lambda_1=\lambda_2$).
The presence of the zero-range perturbation at the point of contact leads to
appearance of virtual levels $\tilde E =E_R-i\Gamma$
in the spectrum of the Hamiltonian $H$.
Positions of the virtual levels are defined by equation
\begin{equation}
\widetilde Q_1(\tilde E)\widetilde Q_2(\tilde E)-|a|^2=0.
\end{equation}
The scattering on the virtual levels leads to appearance of dips on the dependence
$G_{11}(\mu)$.
If the coupling between the wave functions on different cylinders is weak ($|a|\ll 1$),
then the virtual level $\tilde E$ is situated
in the vicinity of the root $\tilde E_1$ of the equation $\widetilde Q_1(\tilde E_1)=0$.
The conductance has only one dip on each plateau in this case.
The dip is situated near the point $\mathop{\rm Re} \widetilde Q_1(E_1)=0$.
If $\lambda_1 \gg r_1$, then the dip is situated near the right edge of
the conductance plateau.
With decreasing $\lambda_1$, the dip shifts to lower energies
and disappears reaching the left edge of the plateau.
There are no dips of the conductance in the limit of the strong
point perturbation ($\lambda_j\ll r_j$) and the weak interaction
between the wave functions ($|a|\ll 1$).
The dependence $G_{11}(\mu)$ is monotonic under these conditions (Fig. \ref{f:G11-2}).

If the coupling between the wave functions
on different tubes is sufficiently strong $|a|\geq 1$,
then additional conductance minima appear on the graph $G_{11}(\mu)$
(Fig.~\ref{f:G11-3}).
These minima are stipulated by splitting of the virtual levels
due to the interaction between electron states on different cylinders.
In the case of the strong point perturbation ($\lambda_j\ll r_j$) and
the strong interaction between the wave functions ($|a|\gg 1$),
the dependence $G_{11}(\mu)$ contains one dip on each conductance plateau.

Let us determine the minimal values of the conductance $G_{11}(\mu)$.
We denote $\xi_j=\mathop{\rm Re} \widetilde Q_j/\mathop{\rm Im} \widetilde Q_j$ and
$\eta=|a|^2/(\mathop{\rm Im} \widetilde Q_1\mathop{\rm Im} \widetilde Q_2)>0$.
Then we can express the conductance $G_{21}(\mu)$
in terms of three real variables $\xi_1$, $\xi_2$ and $\eta$
\begin{equation}
\label{G11f}
G_{11}(\mu)=G_0 \left(N(\mu)-\frac{1}{1+f_{11}(\mu)}\right),
\end{equation}
where
\begin{equation}
\label{f11}
f_{11}(\mu)=\frac{(\xi_1\xi_2-\eta)^2+\xi_1^2}{1+\xi_1^2+2\eta}\geq 0.
\end{equation}
One can see from Eq.~(\ref{G11f}) that the depth of dips equals $G_0$
when $f_{11}=0$. That is possible only when the interaction between the tubes is absent
($a=0$). Otherwise, the depth of dips is less than $G_0$.

If radii of nanotubes are different, then the
dependence $G_{11}(\mu)$ contains additional peaks and dips
in the vicinity of the points $E_m^{(2)}$.
Similar effects occur in the case of different electron effective masses.
To study the form of the curve $G_{11}(\mu)$ we consider the asymptotics
of the Q-function in the vicinity of $E_m^{(j)}$.
One can see from Eq.~(\ref{eq:Qj}) that $\widetilde Q_j(E)$ has a root singularity
at the point $E_m^{(j)}$
\begin{equation}
\label{asympQ}
\widetilde Q_j(E)=
\begin{cases}
2\sqrt{\frac{\varepsilon_j}{E_m^{(j)}-E}}+F(E),& E\to E_m^{(j)}-0, \\
2i\sqrt{\frac{\varepsilon_j}{E-E_m^{(j)}}}+F(E),& E\to E_m^{(j)}+0,
\end{cases}
\end{equation}
where $F(E)$ is a smooth function of $E$.
It is worth mentioning that the root singularity similar to Eq.~(\ref{asympQ})
is present in the electron density of states on the cylinder.
It is convenient to represent $Q_j(E)$ in the form $Q_j(E)=|Q_j(E)|\exp(i\phi_j)$.
One can see from Eq.~(\ref{asympQ}) that the argument
$\phi_j(E)$ changes abruptly by $\pi/2$ at the point $E_m^{(2)}$.
The behavior of the function $\phi_j(E)$
in the interval $(E_m^{(j)},E_{m+1}^{(j)})$ depends
on the scattering length $\lambda_j$.
In the case of the strong point potential ($\lambda_j\ll r$)
the argument $\phi_j(E)$ decreases monotonically from $\pi/2$ to zero
when the energy varies from $E_m^{(j)}$ to $E_{m+1}^{(j)}$.
In the case of the large scattering length ($\lambda_j\gg r$) the argument $\phi_j$
has a maximum $\phi_j^{\rm max}$ in the range ($\pi/2<\phi_j^{\rm max}<\pi$).
Using Eq.~(\ref{asympQ}), we obtain the following asymptotics for
$G_{11}(\mu)$ in the vicinity of  $E_m^{(2)}$:
\begin{equation}
\label{asympG}
G_{11}(\mu)\simeq G_1(\mu)+G_2(\mu),
\end{equation}
where $G_{1}(\mu)$ is given by
\begin{equation}
\label{Greg}
G_{1}=G_0\left(N(\mu)-\frac{(\mathop{\rm Im} \widetilde Q_1)^2}{|\widetilde Q_1|^2}\right),
\end{equation}
and $G_{2}(\mu)$ has the form
\begin{equation}
\label{G2}
G_{2}(\mu)=-\frac{G_0|a|^2}{|\widetilde Q_1|}\sqrt{|\mu-E_m^{(2)}|}\sin 2\phi_1 \sin(\phi_1+\phi_2).
\end{equation}
Here $G_{1}(\mu)$ is the conductance of the single
cylinder with the point perturbation.
This term is regular at the point $\mu=E_m^{(2)}$.
The second term $G_{2}(\mu)$ in Eq.~(\ref{asympG})
is stipulated by the influence of the second tube.
Eq.~(\ref{G2}) shows that the derivative of $G_{2}(\mu)$ has
the root singularity at the point $E_m^{(2)}$.

\begin{figure}[htb]
\includegraphics[width=0.95\linewidth]{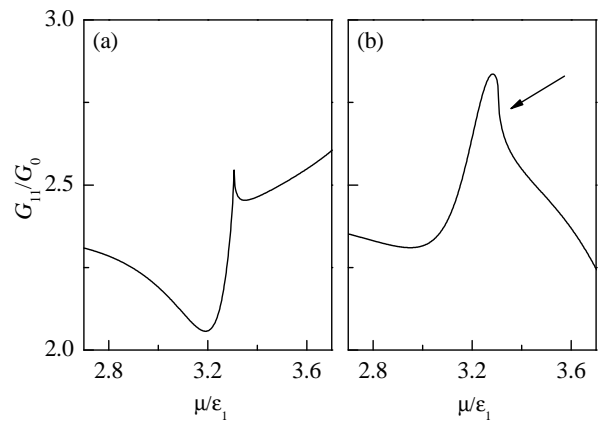}
\caption{\label{f:G11-5}
Conductance  $G_{11}$ versus the Fermi energy $\mu$
at $r_2=1.1 r_1$ and $a=10$.
(a) $\lambda_1=\lambda_2=0.3r_1$ ($\phi_1<\pi/2$);
(b) $\lambda_1=\lambda_2=5r_1$ ($\phi_1>\pi/2$).
The arrow marks the point $\mu=E_m^{(2)}$.}
\end{figure}

Two different line shapes are possible depending on the value of $\phi_1$.
If $\phi_1<\pi/2$, then $\sin(\phi_1+\phi_2)>0$ and $\sin(2\phi_1)>0$
on both sides of $E_m^{(2)}$.
Under these conditions, $G_{2}(\mu)$ is negative and reaches
its maximal value $G_{2}=0$ at the point $E_m^{(2)}$.
Hence, the dependence $G_{11}(\mu)$
has a sharp peak at the point $E_m^{(2)}$ (Fig.~\ref{f:G11-5}a).
If $\phi_1>\pi/2$, then the sign of $G_{2}(\mu)$ is changed at $\mu=E_m^{(2)}$
and the curve $G_{11}(\mu)$ has the form represented in Fig.~\ref{f:G11-5}b.

It is obvious that at a finite temperature the dependence $G_{11}(\mu)$ is smooth
and the asymmetric peaks resemble Fano resonances.
However, the origin of peaks and dips in the conductance of crossed cylinders
differs from the origin of the Fano resonances.
Fano resonances emerge from the coherent interaction of a discrete
state and a continuum while the peaks and dips in the conductance $G_{11}(\mu)$
are stipulated by root singularities in the density of states.

The second term in Eq.~(\ref{G11}) has a step down of amplitude $G_0$
at the point $\mu =E_m$. Hence, the amplitude of the conductance steps
equals $G_0$ in contrast to the case of unperturbed
cylinder where the amplitude equals $2G_0$.

\begin{figure}[htb]
\includegraphics[width=\linewidth]{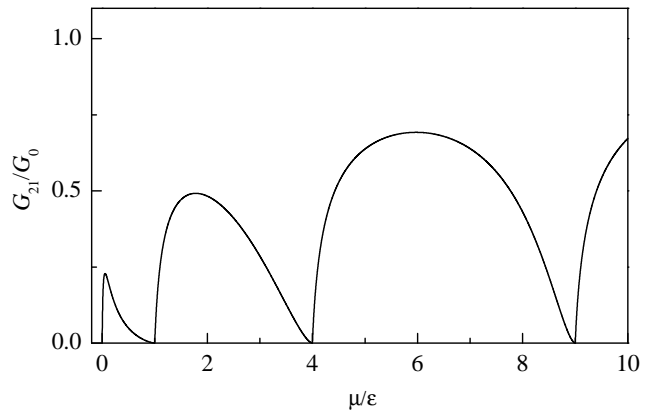}
\caption{\label{f:G21-2}
Conductance $G_{21}$ versus the Fermi energy $\mu$
at $r_1=r_2$, $\lambda_1=\lambda_2=0.1 r_1$, and $a=5$.}
\end{figure}

\begin{figure}[htb]
\includegraphics[width=\linewidth]{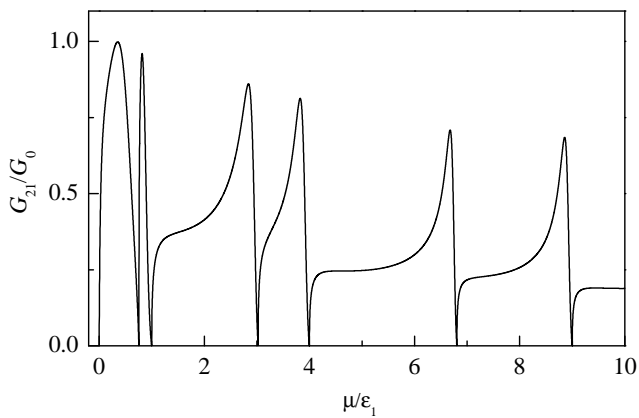}
\caption{\label{f:G21-3}
Conductance $G_{21}$ versus the Fermi energy $\mu$
at $r_2=1.15r_1$, $\lambda_1=\lambda_2=5r_1$, and $a=8$.}
\end{figure}

Let us consider now the conductance $G_{21}$ that corresponds
to the transmission of electrons from the first cylinder to the second one.
Conductance $G_{21}$ determines transport properties of the system in the case
when the contacts $1a$ and $1b$ in Fig.~\ref{scheme} have equal potential $V_1$ while
the contacts $2a$ and $2b$ have the potential $V_2$.
\begin{figure}[htb]
\includegraphics[width=\linewidth]{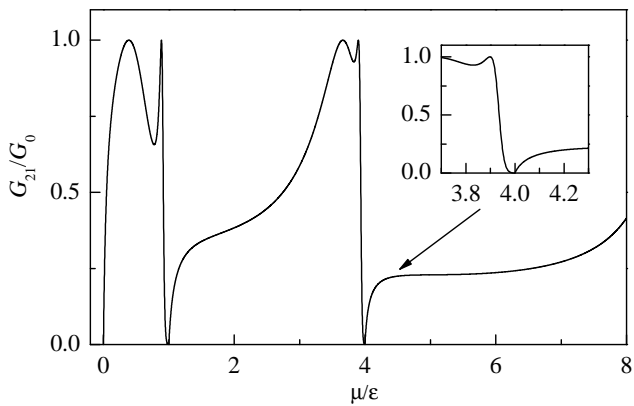}
\caption{\label{f:G21-1}
Conductance $G_{21}$ versus the Fermi energy $\mu$
at $r_1=r_2$, $\lambda_1=\lambda_2=3r_1$, and $a=5$.
The inset represents the same dependence in the vicinity of $\mu=4 \varepsilon $.}
\end{figure}
There are four ways for an electron
to pass from the first tube to the second one:
$1a\to 2a$, $1a\to 2b$, $1b\to 2a$, and $1b\to 2b$.
Due to the symmetry of the system the transmission amplitudes of all
the transitions are equal.
The transmission amplitudes $t_{m'm}^{21}$ are determined in the same way
as $t_{m'm}^{11}$
\begin{equation}
\label{t21}
t_{m'm}^{21}(E)=\frac{i\widetilde Q_2}{\widetilde Q_1\widetilde Q_2-|a|^2}.
\end{equation}
Using the Landauer formula and Eq.~(\ref{eq:ImQj}), we obtain
\begin{equation}
\label{G12}
G_{21}(\mu)=4 G_0 \frac{|a|^2 \mathop{\rm Im} \widetilde Q_1\mathop{\rm Im} \widetilde Q_2}{|\widetilde Q_1 \widetilde Q_2-|a|^2|^2}.
\end{equation}
Here the  factor $4$ is stipulated by four ways for the electron to propagate from
reservoirs marked $1a$ and $1b$ in Fig.~\ref{scheme} to reservoirs marked $2a$ and $2b$.
The dependence $G_{21}(\mu)$ is represented in Figs.~\ref{f:G21-1}-\ref{f:G21-3}.

One can see from Eq.~(\ref{G12}) that the conductance $G_{21}(\mu)$ vanishes when
the Fermi energy $\mu$  coincides with $E_m^{(j)}$.
Using Eq.~(\ref{asympQ}), we get the following form for $G_{21}(\mu)$
in the vicinity of $E_m^{(j)}$:
\begin{equation}
\label{asympG21}
G_{21}(\mu)=
\begin{cases}
A_l(E_m^{(j)}-\mu),& \mu\to E_m^{(j)}-0, \\
A_r\sqrt{\mu-E_m^{(j)}}, & \mu\to E_m^{(j)}+0.
\end{cases}
\end{equation}
where $A_l$ and $A_r$ are positive factors.
One can see that the dependence $G_{21}(\mu)$
exhibits a sharp kink at the point $E_m^{(j)}$ (Fig.~\ref{f:G21-1}).
The kink is stipulated by the root singularity in the density of states.
Similar line shape had been found earlier
in the conductance of the quantum cylinder
with one-dimensional leads. \cite{MP05}
An asymmetric conductance peak appears in a vicinity of the point $\mu=E_m^{(j)}$.
The form of the peak is similar to the Fano resonance line shape,
however the dependence $G_{21}(\mu)$ at $T=0$
is not smooth in contrast to the Fano curve.

Let us determine the maximal value of the conductance $G_{21}(\mu)$.
Using real variables $\xi_j=\mathop{\rm Re} \widetilde Q_j/\mathop{\rm Im} \widetilde Q_j$
and $\eta=|a|^2/(\mathop{\rm Im} \widetilde Q_1\mathop{\rm Im} \widetilde Q_2)$,
we can express the conductance $G_{21}(\mu)$ in the following form:
\begin{equation}
G_{21}(\xi_1,\xi_2,\eta)=
4 G_0 \frac{\eta}{(\xi_1\xi_2-\eta-1)^2+(\xi_1+\xi_2)^2}.
\end{equation}
If $\eta < 1$, then $G_{21}$ as a function of $\xi_1$ and $\xi_2$ has
only one maximum $G_{21}=4G_0\eta/(\eta+1)^2$ at $\xi_1=\xi_2=0$.
It is obvious, that $G_{21}<G_0$ in this case.
The condition $\xi_1=\xi_2=0$ is equivalent to
$\mathop{\rm Re} Q_1=\mathop{\rm Re}Q_2=0$.
That is possible only in the case of identical tubes and large scattering lengths.
In the opposite case of strong point perturbation ($\lambda_j\ll r_j$) the function
$\mathop{\rm Re} Q_j(E)$ has no zeros and the maxima of the conductance
are situated near the minima of $\mathop{\rm Re} Q_j(E)$ (Fig.~\ref{f:G21-3}).
In this limit, the maximal value of the conductance is smaller than $G_0$.

An additional maximum $G_{21}=G_0$ appears at $\xi_1=\xi_2=\sqrt{\eta-1}$
in the case $\eta \geq 1$ (Fig.~\ref{f:G21-1}).
That means the coupling between wave functions
on different nanotubes is strong.
The conductance $G_{21}$ decreases with decreasing of the coupling.
Therefore, the conductance $G_{21}$ never exceeds $G_0$.
We relate this result to limited transparency of the point contact.
We stress that the result is not trivial because
the sum in Landauer's equation (\ref{Land}) contains many terms.
The condition $\xi_1=\xi_2$ is satisfied only if the tubes are identical.
Hence, the conductance $G_{21}$ reaches its maximal possible value only in the
case of identical tubes and strong coupling
between the wave functions ($|a|\gtrsim 1$).

If radii of tubes are different, then the number of zeros
on the dependence $G_{21}(\mu)$ increases
because $E_{m}^{(1)}\neq E_{m}^{(2)}$ (Fig.~\ref{f:G21-3}).
The maximal value of the conductance $G_{21}$
is also smaller then $G_0$ in this case.

\section{Conclusion}

The electron transport in crossed conductive nanocylinders is investigated
using the Landauer--B\"uttiker formalism.
An explicit form for the conductance $G_{11}$ as a function of the Fermi energy
$\mu$ is obtained.
We have shown that the conductance of each tube contains
resonance dips stipulated by the back-scattering of electrons
on the contact.
The maximal value of dips does not exceed a unit of the conductance quantum.
Positions of dips depend on parameters of the contact.
In the case of identical tubes the conductance can exhibit
no more than two dips on each plateau.
If the radii of tubes are different, then the dependence
of the conductance on the Fermi energy contains asymmetric
peak-dip structures of the form given by Eq.~(\ref{asymp}).
The dips and peaks are stipulated
by the root singularities in the density of states.
It is worth mentioning that
the asymmetric peaks and dips have been observed experimentally in the
conductance of many carbon nanotube based systems,
in particular in crossed nanotubes. \cite{Kim03}.
The asymmetric line shapes were attributed to Fano resonances that
emerge from the interference of the bound states with the continuum.
Results of the present paper provide alternative possible explanation
of the asymmetric peaks and dips. According to our results
they might originate from the root singularities in the density of states.

The conductance $G_{21}$ which is related to transmission
of electrons from the first cylinder to the second one
has the resonance nature.
The maximal value of the conductance $G_{21}$ is
a unit of the conductance quantum.
The condition of maximal transmission is satisfied only if
the tubes are identical.
Similar results have been obtained in Ref.~\onlinecite{Dag04} for several
crossed junctions of small nanotubes.
It should be noted, that the resistance of the value 16.8~k$\Omega \gtrsim G_0^{-1}$
has been observed experimentally \cite{Kim03} for crossed carbon nanotubes at $T=4.2$~K.

The work is support by the Russian Foundation for Basic Research
(grant No. 05-02-16145).

\end{document}